# Point Defects at Cleaved $Sr_{n+1}Ru_nO_{3n+1}$ (001) Surfaces

Bernhard Stöger[1], Marcel Hieckel[1,2], Florian Mittendorfer[1,2], Zhiming Wang[1], Michael Schmid[1], Gareth S. Parkinson[1], David Fobes[3], Jin Peng[3], John. E. Ortmann[3], Andreas Limbeck[4], Zhiqiang Mao[3], Josef Redinger[1,2], Ulrike Diebold[1]

[1]Institute of Applied Physics, Vienna University of Technology, Wiedner Hauptstraße 8-10, A-1040 Vienna, Austria

[2]Center for Computational Materials Science, Vienna University of Technology, Gußhausstraße 25-25a, A-1040 Vienna, Austria

[3]Department of Physics and Engineering Physics, Tulane University, New Orleans, Louisiana 70118, USA

[4]Institute of Chemical Technologies and Analytics, Vienna University of Technology, Getreidemarkt 9, A-1060 Vienna, Austria

The (001) surfaces of cleaved $Sr_3Ru_2O_7$ and $Sr_2RuO_4$ samples were investigated using low-temperature scanning tunneling microscopy and density functional theory calculations. Intrinsic defects are not created during cleaving. This experimental observation is consistent with calculations, where the formation energy for a Sr and O vacancy, 4.19 eV and 3.81 eV respectively, is significantly larger than that required to cleave the crystal, 1.11 eV/(1x1) unit cell. Surface oxygen vacancies can be created through electron bombardment, however, and their appearance is shown to vary strongly with the imaging conditions. Point defects observed on as-cleaved surfaces result from bulk impurities and adsorption from the residual gas.

**Introduction**

The Ruddlesden-Popper series $A_{n+1}B_nO_{3n+1}$ of perovskite-type materials exhibit a wide range of distinct electronic and magnetic properties, including unconventional superconductivity, Mott insulating state, itinerant magnetism and colossal magnetoresistance. Due to strong coupling between charge, spin, lattice and orbital degrees of freedom [1], physical properties of the perovskite-type materials are extremely sensitive to perturbations such as small structural changes, and can be tuned via non-thermal parameters such as magnetic fields, pressure, or chemical doping. The strontium ruthenate series ($Sr_{n+1}Ru_nO_{3n+1}$), for example, ranges from a spin-triplet superconductor (*n*=1, $Sr_2RuO_4$) [2,3], over an itinerant metamagnetic state with a field-tuned nematic phase ($n=2$, $Sr_3Ru_2O_7$) [4–6], to an itinerant ferromagnet ($n=\infty$, $SrRuO_3$) [7,8].

Substitution of Ca for Sr also affects the physical properties drastically by distorting the crystal structure. The $Ca_{2-x}Sr_xRuO_4$ system evolves from spin-triplet superconductivity to a nearly ferromagnetic state for $x = 0.5$ and then finally to a Mott insulating state for $x<0.2$ [9], while in $(Sr_{1-x}Ca_x)_3Ru_2O_7$ the system varies from an itinerant metamagnetic state to a cluster spin glass state and finally to an antiferromagnetic state consisting of ferromagnetic bilayers when Sr is gradually replaced by Ca [10–12]. Moreover, dopants and oxygen vacancies often act as reactive sites on metal oxides, where preferred adsorption or dissociation of molecules takes place, which makes them especially important for catalytic applications [13]. For example, we have shown recently that CO adsorbs first at Ca dopants on $Sr_3Ru_2O_7$, before adsorbing on the clean surface [14].

Because of the layered nature of these materials, surface probes such as Scanning Tunneling Microscopy (STM) are ideally suited for obtaining direct information about local structural, stoichiometric and electronic inhomogeneities. Extrinsic defects (i.e., dopants) can have drastic effects, as has been shown in studies of, e.g., lightly Ti-doped $Sr_2RuO_4$ [15] and $Sr_3Ru_2O_7$ [16] as well as Mn-doped $Sr_3Ru_2O_7$ [17]. These studies established that Sr atoms are imaged as bright protrusions and oxygen atoms as dark depressions in STM images; that dopant atoms in the surface layers can be distinguished from the regular lattice; and that Mn dopants affect the distortion of the $RuO_6$ octahedra of $Sr_3Ru_2O_7$ [17]. Lee et al. [16] show that Ti defects act as scattering centers leading to quasiparticle interferences on the $Sr_3Ru_2O_7$ surface. Even without doping, the as-cleaved surfaces themselves can exhibit interesting properties. Matzdorf et al. [18,19] demonstrated that cleaving $Sr_2RuO_4$ along the [001] direction creates flat and well-ordered surfaces. They observed a c(2×2) reconstruction by STM and LEED, and the accompanying lattice distortion was attributed to a ferromagnetic ground state at the surface. Subsequent studies on cleaved $Sr_2RuO_4$ [20] and $Sr_3Ir_2O_7$ [21] surfaces have proposed that intrinsic defects in the surface lattice are formed during cleaving. Pennec et al. [20] suggested that this process is temperature driven, and that higher temperatures lead to higher concentrations of intrinsic defects.

These latter studies have motivated us to systematically investigate surface defects that are present on as-cleaved surfaces. Using low-temperature scanning tunneling microscopy (LT-STM) and density functional theory (DFT) calculations we have investigated a total of thirty cleaved $Sr_3Ru_2O_7$, $Sr_2RuO_4$, $(Sr_{0.95}Ca_{0.05})_3Ru_2O_7$, $Sr_2Ru_{0.97}Mn_{0.03}O_4$ and $Sr_2Ru_{0.97}Co_{0.03}O_4$ single crystal surfaces. We demonstrate that no intrinsic defects (Sr / O vacancies) form as a result of the cleaving process, irrespective of the cleaving temperature. Rather, the observed point defects are attributed to bulk impurities and to adsorption from the residual gas.

**Experimental Procedure**

The experiments were carried out in a two-chamber UHV-system with base pressures of $2\times10^{-11}$ and $6\times10^{-12}$ mbar in the preparation and STM chamber, respectively. A low-temperature STM (Omicron LT-STM) was operated at 78 K in constant-current mode using electro-chemically etched W-tips. Positive (negative) sample bias results in STM images of the unoccupied (occupied) states. The samples used in this study were grown by the floating zone technique using a two-mirror optical floating-zone furnace. A detailed description of the growth procedure is found in Ref. [22]. Over the course of the experiments we studied 15 $Sr_3Ru_2O_7$ samples from three different batches, two $(Sr_{0.95}Ca_{0.05})_3Ru_2O_7$ samples, five $Sr_2RuO_4$ samples, one $Sr_2Ru_{0.97}Mn_{0.03}O_4$ sample, and one $Sr_2Ru_{0.97}Co_{0.03}O_4$ sample. The samples were fixed on Omicron sample plates using conducting silver epoxy glue (EPO-TEK H21D), and a metal stub was glued on top with another epoxy adhesive (EPO-TEK H77). The crystals were cleaved by applying a tangential force to the metal stub with a wobble stick. Cleaving was performed in the analysis chamber at 105 K and 300 K. Each crystal could be cleaved several times, after gluing a new stub to the remainder of the previous cleavage. After cleaving, the sample was immediately transferred into the cold STM; first images were usually obtained within 30 min. Electron bombardment was performed by a well-outgassed electron gun in the preparation chamber with the sample held at 105 K.

The composition of the samples was determined via inductively coupled plasma mass spectroscopy (ICP-MS) using laser ablation (LA) for direct analysis of the solid samples. The signal from Ca isotopes was overlapping with doubly charged Sr isotopes in ICP-MS, thus inductively coupled plasma optical emission spectroscopy (ICP-OES) was used to quantify Ca.

DFT calculations were performed with the Vienna Ab-initio Simulations Package (VASP) [23] in the PAW framework [24], using the Perdew Burke Ernzerhof (PBE) exchange-correlation functional [25]. Theoretical studies on $Sr_3Ru_2O_7$ compounds using DFT and approaches beyond [4,26–29], indicate that, despite of the strongly correlated nature of these materials, a reasonable description of the electronic structure can already be obtained on a DFT level [4,28,29]. For the determination of the rotation and tilting of the $RuO_6$ octahedra at the bare surfaces up to three $c(2\times2)$ $Sr_3Ru_2O_7$ perovskite bilayers with a primitive lateral lattice constant of 3.932 Å were used. The investigation of the defects was performed using a single $Sr_3Ru_2O_7$ perovskite bilayer in a larger $(4\times4)$ supercell to reduce spurious interactions between the defects. In the latter case, the uppermost three atomic layers of the perovskite bilayer were fully relaxed performing the Brillouin zone integration on a $2\times2\times1$ Monkhorst-Pack k-point mesh. The vacancy energies are given with respect to molecular oxygen and bulk (fcc) strontium.

## Results

The crystal structure of $Sr_3Ru_2O_7$ is shown in Fig. 1(a). With a value of 1.11 eV/(1×1) unit cell (0.07 eV/Å²), the cleaving energy is lowest for the SrO rock salt structure between the $SrRuO_3$ layers. This is in agreement with previous studies on $Sr_2RuO_4$, $Sr_3Ru_2O_7$ and $Sr_3Ir_2O_7$ [13, 14, 18, 21], which also suggest that the weakest bonds are in the rock salt structure. A $RuO_2$ termination as reported in Ref. [30] has never been observed. Fig. 1(b) shows a top view of the $Sr_3Ru_2O_7$ (001) surface that results from cleaving. Our calculations predict a rotation of the $RuO_6$ octahedra by an angle of ~11° at the surface and ~9° in the bulk. This alternating clockwise and counter clockwise rotation of neighboring $RuO_6$ octahedra is clearly visible in Fig. 1(b), and results in the so-called c(2×2) symmetry observed in low energy electron diffraction [16–19,21]. (We note that, for $Sr_3Ru_2O_7$, the surface layer has the same symmetry as a bulk truncation. Nevertheless, it is standard notation to use the tetragonal unit cell [see Fig. 1(a), yellow square], where the octahedra are not rotated, as the (1x1) unit cell and the primitive unit cell of $Sr_3Ru_2O_7$ is described as a c(2×2) structure [see Fig. 1(a), black square].) The surface layer is reported to exhibit additional distortions compared to a bulk truncated structure, with a more pronounced rotation and a slight tilt of the octahedra with respect to the surface normal [31]. It should be noted that our DFT calculations predict that an additional tilting of the surface octahedra is energetically unfavorable. The source of this discrepancy deserves further investigation, but we note that additional adsorbates at the surface, such as CO, can also induce pronounced structural distortions [14].

Large scale STM images [see Fig. 1(c)] of the cleaved $Sr_3Ru_2O_7$ (001) surface show terraces of several μm² in size separated by steps with a height of 1.1 nm (or multiples thereof), consistent with the bulk periodicity, i.e., the distance between the rock salt layers. Occasional line defects are observed. An analysis of the areas separated by these line defects shows that they appear as antiphase domain boundaries of the c(2×2) rotation of the octahedra at the surface; likely these are manifestations of twin boundaries in the bulk.

Figure 2 shows atomically resolved STM images of three $Sr_3Ru_2O_7$ (001) surfaces following cleaving at either 105 K or 300 K. Sr and O atoms of the SrO layer are imaged as bright protrusions and dark depressions in STM, respectively. The surface of sample #6 [see Fig. 2(a)] exhibits a low concentration of surface defects. The primary defects observed are large, dark cross features related to CO molecules, which readily adsorb from the residual gas, and have been described in detail in Ref. [14]. No additional point defects are observed that might be attributable to either O or Sr vacancies. This is consistent with our DFT calculations, which predict that such defects cost significantly more energy than that required to cleave the sample (1.11 eV/(1×1) unit cell, i.e. 0.56 eV per broken Sr-O bond). Specifically, 4.19 eV is necessary to create a Sr vacancy and 3.81 eV is required to create an O vacancy. Increasing the cleaving temperature to 300 K, which corresponds to an additional thermal energy of 20 meV, should therefore have little effect. STM images acquired following 300 K cleaving [see Fig. 2(b)] are indeed similar to those acquired at 105 K, with the addition of bright features. Purposely dosing $H_2O$ on a cold surface shows that these bright protrusions (often arranged as monomers, dimers, and one-dimensional lines) result from water adsorption [see inset Fig. 2(b)]. The details of water-related structures and their formation will be reported in a forthcoming publication [32].

On many oxides, surface oxygen vacancies ($V_O$s) can be created by electron bombardment [33,34]. Following 45 minutes of 1000 eV electron bombardment at 105 K (current: ~5 µA, bombarded area: ~1 mm²), new features appear that are centered at the position of the apical oxygen/subsurface Ru atom [see Fig. 3]. Counting the number of these features in the STM images yields an estimated cross section for electron induced oxygen desorption of the order of ~$10^{-20}$ cm². In empty-states images [see Fig. 3(a)] $V_O$s appear as small, dark crosses, while in filled states they appear as bright squares [see Fig. 3(b)]. Such features were not observed on the as-cleaved surface. While, at first sight, these cross-like features may look similar to those identified as adsorbed CO molecules [14], CO and oxygen vacancies can, in fact, be easily distinguished. Adsorbed CO appears significantly larger and show a 2-fold symmetry, reflecting the carboxylate that forms when the molecule reacts with a lattice O [14]. In contrast, the oxygen vacancies show 4-fold symmetry [see Fig. 3(a), inset]. In addition, CO can easily be removed by scanning with the STM tip at a positive sample bias voltages of at least 2.7 V, as was shown in our previous work [14]. Simulated STM images of O vacancies show reasonable agreement for occupied states, but the calculations also predict a bright appearance for the unoccupied states. This might be related to an underlocalization of the vacancy electrons on the DFT/PBE level, which can result in an overestimation of the corresponding band width, and an additional error in the band position.

STM images obtained from sample #5 [see Fig. 2(c)] exhibit a much higher concentration of point defects, approximately 0.008 monolayer (where 1 monolayer (ML) is defined as the amount of Sr and O atoms in the surface layer). Although the sample in Fig. 2(a) is from the same sample batch as the sample in Fig. 2(c) the defect coverage varies in these two figures. A macroscopic change of the scanned area, or cleaving the same crystal several times, revealed that the number of defects can vary significantly on one and the same crystal from sample batch #2. A trace analysis revealed that sample #5 contains Mg, Al, Cr, Fe and Ni impurities and sample #8 Ca impurities on the order of a few hundred ppm [see Table. 1]. This corresponds to a defect concentration of ~0.006 ML, in agreement with the density observed in Fig. 2(c).

To investigate the appearance of impurity atoms in STM, we resort to the surfaces of intentionally-doped $Sr_3Ru_2O_7$ single crystals. Figure 4(a) shows an image of $(Sr_{0.95}Ca_{0.05})_3Ru_2O_7$ following cleaving at 105 K. Black [see Fig. 4(a), red arrow] and grey [see Fig. 4(a), blue arrow] defects, roughly equal in number, are located at the position of surface Sr atoms. The depressions have 2-fold, not 4-fold symmetry; they are elongated along either the [100] or [010] direction as seen in Fig. 2(c), depending on the position within the c(2×2) unit cell. We assign the black and gray defects to Ca dopants substituted for Sr atoms in the surface and subsurface layer, respectively. An STM simulation of Ca in the surface [see inset Fig. 4(a)] confirms its dark appearance. This change in the contrast is predominantly due to a geometric effect, as the Ca impurity is positioned 0.19 Å below the surrounding Sr ions. STM images of cleaved $Sr_2Ru_{0.97}Mn_{0.03}O_4$ and $Sr_2Ru_{0.97}Co_{0.03}O_4$ [Figs. 4(b, c)] samples reveal bright defects at the apical oxygen position in filled- and empty-states STM images.

**Discussion**

Our combination of experimental and theoretical data demonstrate that the point defects observed on cleaved strontium ruthenate samples result from adsorption and bulk impurities, and not from intrinsic point defects (O or Sr vacancies) created during the cleaving process. This is in contrast to the reports by Pennec et al. [20] on $Sr_2RuO_4$ and Okada et al. [21] on $Sr_3Ir_2O_7$, who proposed vacancies to form as a result of cleaving. Given the similarity between Ruddlesden Popper materials, it is unlikely that the energy required to form an oxygen vacancy in the surface SrO layer differs markedly from $Sr_3Ru_2O_7$, or that the cleaving energy at the rock salt layer is much larger. Our STM images of surface oxygen vacancies, purposely created by electron bombardment, show that the appearance of these features depends strongly on the tunneling conditions. Furthermore, we find no evidence of a deterioration of the sample surface with cleaving temperature. Consequently, the features reported in Refs. [20,21], are likely not intrinsic defects, but result from adsorption and/or bulk impurities.

Our results clearly show that bulk impurities in 100 ppm range already lead to visible point defects, and that these atoms can appear as both protrusions and depressions at the Sr and O lattice sites, depending on the atom type and configuration. The growth of ultrahigh-purity strontium ruthenates is not easily achieved since many metals can substitute into the lattice of the starting materials, i.e. RuO and $SrCO_3$, do not have ultra-high purity. Even small amounts of impurities show up prominently in the STM images. So far, Ti dopants in $Sr_2RuO_4$ [15] and $Sr_3Ru_2O_7$ [16], as well as Mn-doped $Sr_3Ru_2O_7$ [17] single crystals have been identified by STM. Our work adds Ca-doped $Sr_3Ru_2O_7$, Mn- and Co-doped $Sr_2RuO_4$ to this database. Our studies have also revealed that the SrO-terminated surface of these materials is highly reactive toward the molecules present in a typical vacuum system [14]. Both water and CO are found to adsorb readily at regular lattice sites, suggesting that the adsorbate coverage would increase rapidly in less pristine vacuum conditions. As adsorbates usually cause relaxations in the lattice, the resulting structural distortions can also give rise to change in the local electronic structure in these strongly-correlated materials.

Adsorption from the residual gas may explain puzzling ARPES results, [35,36] in which $Sr_2RuO_4$ samples cleaved at 180 K exhibited suppressed surface states linked to the c(2×2) surface reconstruction [18,19], a phenomenon not observed in Ref. [37]. Our study shows no discernible difference in the quality of the surface obtained at different cleaving temperatures, in agreement with Ref. [38]. However, in poor vacuum conditions a difference could occur if cold surfaces in the vicinity of the sample act as a cryogenic pump, reducing the local pressure around a freshly cleaved surface. Cleaving at room temperature would have no such benefits, and the sample would suffer more contamination. Such considerations highlight the difficulty in linking changes in electronic structure observed in ARPES to modifications of the atomic structure in the absence of atomic-scale imaging methods. Our electron bombardment results show that creating O vacancies by electronic excitations is straightforward on strontium ruthenate surfaces. X-rays can also induce desorption via electronic transitions, thus possible beam damage needs to be considered in synchrotron experiments. Recently Santander-Syro et al. [39] initially attributed a two-dimensional electron gas formed at cleaved $SrTiO_3$ surfaces to oxygen vacancies created during cleaving, but Meevasana et al. and Wang et al. [40,41] subsequently showed that the vacancies result from photon irradiation.

In summary we showed that clean, defect-free $Sr_3Ru_2O_7$(001) surfaces can be created by cleaving in UHV, independent of the cleaving temperature. No intrinsic defects are formed during the cleaving process, but the surface is reactive toward adsorption of molecules from the residual gas. Previous reports of surface vacancies on cleaved surfaces likely results from bulk impurities.

Acknowledgement. This work was supported by the ERC Advanced Research Grant "OxideSurfaces" and by the Austrian Science Fund (FWF Project No. F45). The calculations have been performed on the Vienna Scientific Cluster (VSC-2). The work at Tulane is supported by the NSF under grant DMR-1205469. The crystals structures were plotted using VESTA [42]. VISIT [43] was used to visualize the simulated STM data.

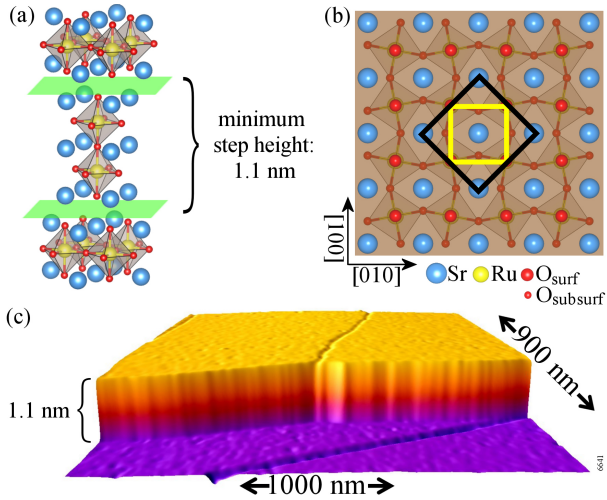

**FIG. 1**. (Color online) (a) The crystal structure of $Sr_3Ru_2O_7$. DFT calculations predict a cleaving energy of 1.11 eV/(1×1) unit cell. (b) Top view of the $Sr_3Ru_2O_7$(001) surface, which contains Sr (blue) and O (red, large) atoms only. Smaller red circles indicate second-layer O atoms. The inner (yellow) square marks the tetragonal unit cell; the black square marks the primitive unit cell (conventionally called c(2×2)), which is caused by the clockwise and counterclockwise rotation of the $RuO_6$ octahedra. (c) Large-scale STM image (1000×900 nm², $U_{sample}$ = +1.5 V, $I_t$ = 0.15 nA, $T$ = 78 K).

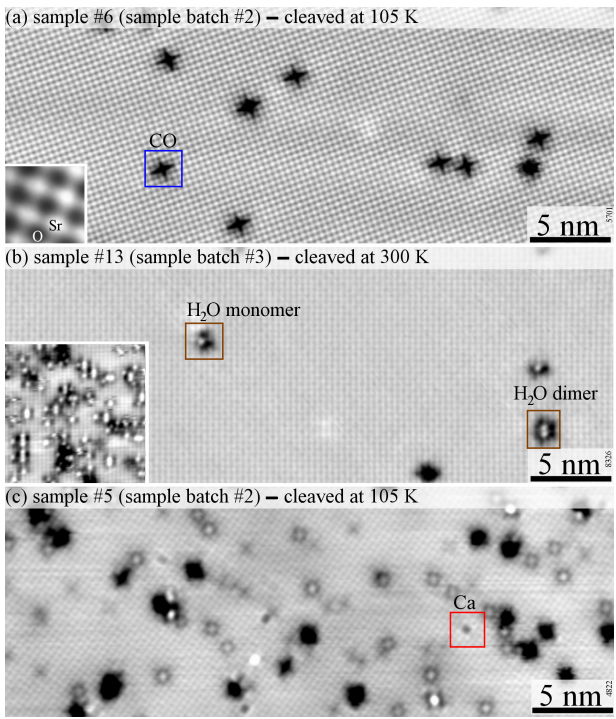

**FIG. 2.** (Color online) STM images of cleaved $Sr_3Ru_2O_7$(001) surfaces (39.5×15.5 nm², $T$ = 78 K). (a) $U_{sample}$: +0.05V, $I_t$: 0.15 nA. Following cleaving at 105 K, sample #6 is free from defects other than adsorbed CO (blue box) molecules identified in Ref. [28]. (b) $U_{sample}$: +0.2 V, $I_t$: 0.15 nA. Following cleaving at room temperature, sample #13 is similarly

defect-free, although some adsorbed water molecules are observed (brown box). After dosing $H_2O$ on a clean surface, the same features can be observed (see inset). (c) $U_{sample}$: +0.05V, $I_t$: 0.15 nA. Many point defects are observed on upon cleaving a different $Sr_3Ru_2O_7$ crystal (cleaving temperature 105 K). These defects occur either at the Sr or the apical oxygen / Ru lattice sites and are related to bulk impurities present in this crystal [see Table 1].

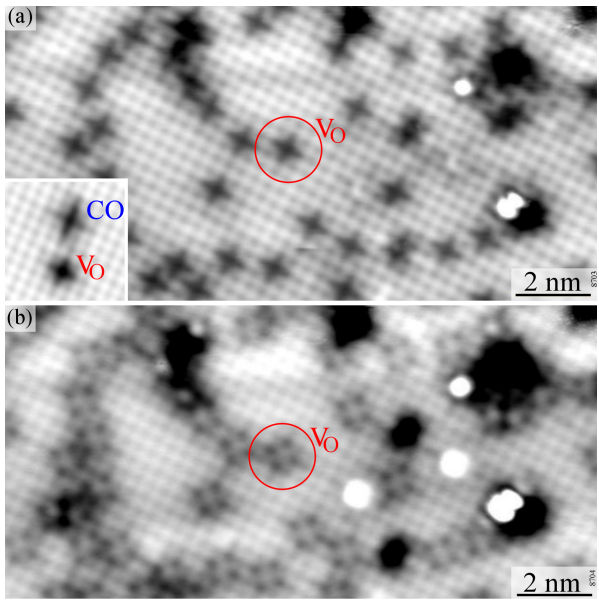

**FIG. 3**. (Color online) STM images of a $Sr_3Ru_2O_7$ (001) surface following bombardment with 1000 eV electrons at 105 K. (a) In empty-states images ($U_{sample}$ = +0.1 V, $I_t$ = 0.15 nA, $T$ = 78 K) oxygen vacancies appear as small dark crosses. They can be easily distinguished from adsorbed CO molecules (see inset). (b) In filled-states images ($U_{sample}$ = -0.4 V, $I_t$ = 0.15 nA, $T$ = 78 K) oxygen vacancies appear as small bright squares. Both features are centered at apical oxygen/subsurface Ru atoms.

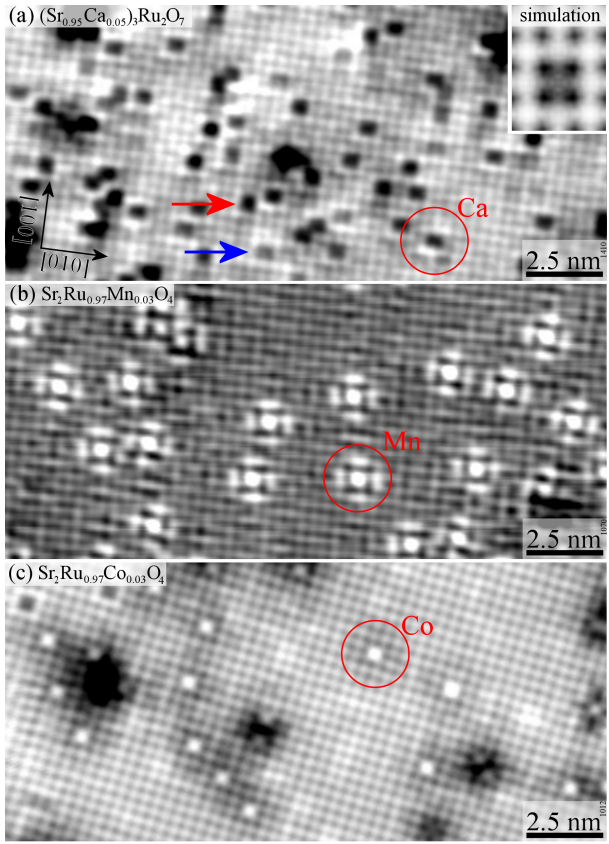

**FIG. 4.** (Color online) STM images of doped $Sr_{n+1}Ru_nO_{3n+1}$ samples. (a) STM image of $(Sr_{0.95}Ca_{0.05})_3Ru_2O_7$ ($U_{sample}$ = +0.3 V, $I_t$ = 0.15 nA, $T$ = 78 K). The Ca atoms replace Sr atoms. Ca atoms in the first and second-layer are imaged as dark and grey squares, respectively. One surface and one subsurface Ca dopant is marked with a red and blue arrow, respectively. An STM simulation confirms the dark appearance of a Ca dopant in the first layer (see inset). (b) STM image of $Sr_2Ru_{0.97}Mn_{0.03}O_4$ ($U_{sample}$ = -0.01 V, $I_t$ = 0.25 nA, $T$ = 6 K), the Mn dopants replace the Ru atoms and appear as crosses with bright spots positioned at the Ru/apical oxygen lattice site. (c) STM topography of $Sr_2Ru_{0.97}Co_{0.03}O_4$ ($U_{sample}$ = +0.05 V, $I_t$ = 0.15 nA, $T$ = 78 K), the Co dopants replace the Ru atoms and appear as bright protrusions positioned at the subsurface Ru /apical oxygen lattice site.

| LA-ICP-MS | Sample #5 / batch #2(ppm) |
|---|---|
| Mg | 167 ± 43.8 |
| Al | 1093 ± 47 |
| Cr | 620 ± 124 |
| Fe | 464 ± 93 |
| Ni | 392 ± 78 |
| LA-ICP-OEM | Sample #8 / batch #2(ppm) |
| Ca | 600 ± 100 |

**TABLE 1.** ICP-MS and ICP-OEM measurements of samples from the same sample batch containing a high level of impurities [see Fig. 2(c)].